\documentclass[twocolumn,english,pra,aps,superscriptaddress]{revtex4-1}
\usepackage[T1]{fontenc}
\usepackage{color}
\usepackage{babel}
\usepackage{mathtools}
\usepackage{bm}
\usepackage{graphicx}
\usepackage{esint}
\usepackage{amsfonts}
\usepackage{graphicx}
\usepackage{float}
\usepackage{subfigure}
\usepackage{amsthm}
\usepackage{amssymb}
\usepackage{physics}

\usepackage[usenames,divipsnames,svgnames,table]{xcolor}
\usepackage[unicode=true,pdfusetitle,
bookmarks=true,bookmarksnumbered=false,bookmarksopen=false,
breaklinks=true,pdfborder={0 0 0},backref=false,colorlinks=true]{hyperref}
\hypersetup{linkcolor=NavyBlue,urlcolor=NavyBlue,citecolor=NavyBlue}
\usepackage{breakurl}

\begin{document}
\title{Lee-Yang zeros and quantum Fisher information matrix in a nonlinear system}

\author{Hong Tao}
\affiliation{Key Laboratory of Optical Field Manipulation of Zhejiang Province
and Department of Physics, Zhejiang Sci-Tech University, Hangzhou 310018, China}
\affiliation{National Precise Gravity Measurement Facility,
MOE Key Laboratory of Fundamental Physical Quantities Measurement, School of Physics,
Huazhong University of Science and Technology, Wuhan 430074, China}

\author{Yuguo Su}
\affiliation{State Key Laboratory of Magnetic Resonance and Atomic and Molecular Physics,
Wuhan Institute of Physics and Mathematics, APM, Chinese Academy of Sciences, Wuhan 430071, China}

\author{Xingyu Zhang}
\affiliation{Department of Physics, Zhejiang University, Hangzhou 310027, China}

\author{Jing Liu}
\email{liujingphys@hust.edu.cn}
\affiliation{National Precise Gravity Measurement Facility,
MOE Key Laboratory of Fundamental Physical Quantities Measurement, School of Physics,
Huazhong University of Science and Technology, Wuhan 430074, China}

\author{Xiaoguang Wang}
\email{xgwang@zstu.edu.cn}
\affiliation{Key Laboratory of Optical Field Manipulation of Zhejiang Province
and Department of Physics, Zhejiang Sci-Tech University, Hangzhou 310018, China}

\begin{abstract}
The distribution of Lee-Yang zeros not only matters in thermodynamics and quantum mechanics, but also in mathematics. Hereby we
propose a nonlinear quantum toy model and discuss the distribution of corresponding Lee-Yang zeros. Utilizing the coupling
between a probe qubit and the nonlinear system, all Lee-Yang zeros can be detected in the dynamics of the probe qubit by tuning
the coupling strength and linear coefficient of the nonlinear system. Moreover, the analytical expression of the quantum Fisher
information matrix at the Lee-Yang zeros is provided, and an interesting phenomenon is discovered. Both the coupling strength and
temperature can simultaneously attain their precision limits at the Lee-Yang zeros. However, the probe qubit cannot work as a
thermometer at a Lee-Yang zero if it sits on the unit circle.
\end{abstract}

\maketitle

\section{introduction}

The Lee-Yang zero is an interesting concept in thermodynamics, which was first proposed and discussed by Lee and Yang
in 1952~\cite{Lee1952,Yang1952}. In the study of the lattice gas and Ising model, Lee and Yang wrote the partition function $Z$
into a polynomial form, i.e., $Z=\sum_n p_n z^n$, and extending $z$ to the complex plane via the analytic continuation, the roots
of the equation $Z=0$ are always distributed on the unit circle. This theorem and the roots are usually referred to as the
Lee-Yang unit circle theorem and Lee-Yang zeros nowadays. The Lee-Yang theorem and zeros have been widely studied in many fields,
such as the field theory~\cite{Simon1973,Fisher1978,Kardar2007}, condensed matterphysics~\cite{Kortman1971,Suzuki1971,Lieb1981,
Monroe1991,Garcia2015,Wei2014,Binek1998,Kim2004,Tong2006,Brandner2017,Lebowitz2012,Frohlich2012,Kist2021,Arndt2000,Vecsei2022},
stochastic processes~\cite{Yoshida2022,Flindt2013,Deger2018,Deger2019,Deger2020} and even pure mathematics~\cite{Nishimori1983,
David2010,Hou2023}. In 2012, Wei and Liu proposed a remarkable scheme for the observation of Lee-Yang zeros via the dynamics of a
probe qubit~\cite{Wei2012}, which is then experimentally realized by Peng \emph{et al.}~\cite{Peng2015} in 2015. In 2019, Kuzmak
and Tkachuk used a similar scheme to study the detection of Lee-Yang zeros of a high-spin system~\cite{Kuzmak2019}. Moreover, the
behaviors of quantum resources like spin squeezing and concurrence at the points of Lee-Yang zeros have also been investigated
recently~\cite{Su2020}.

The quantum Fisher information matrix is another fundamental quantity in quantum mechanics~\cite{Helstrom1976,Holevo1982,Liu2020,
Safranek2018}. It was first provided by Helstrom in the field of quantum parameter estimation, which is the extension of parameter 
estimation in quantum mechanics. In quantum parameter estimation, the quantum Fisher information matrix is the lower bound of the 
covariance matrix for a set of unknown parameters. Denote the covariance matrix as $\mathrm{cov}(\vec{x},\{\Pi_i\})$ with $\vec{x}$ the 
vector of unknown parameters and $\{\Pi_i\}$ a set of positive operator-valued measure, then $\mathrm{cov}(\vec{x},\{\Pi_i\})$ satisfies
the inequality $\mathrm{cov}(\vec{x},\{\Pi_i\})\geq \mathcal{F}$~\cite{Helstrom1976,Holevo1982}, where $\mathcal{F}$ is the quantum
Fisher information matrix for $\vec{x}$. The entry of $\mathcal{F}$ can be calculated via the equation $\mathcal{F}_{ij}=\mathrm{Tr}
(\rho \{L_i,L_j\})/2$ with $L_{i(j)}$ the symmetric logarithmic derivative for the unknown parameter $x_{i(j)}$, $\rho$ the density
matrix and $\{\cdot,\cdot\}$ the anti-commutator. $L_i$ satisfies the equation $\partial_{x_i}\rho=(\rho L_i+L_i\rho)/2$. Nowadays,
the quantum Fisher information matrix has been widely considered as a fundamental quantity in quantum mechanics due to its good
mathematical properties and wide connections to other aspects of quantum mechanics.

It is known that the long-range Ising model can be mapped into the generalized one-axis twisting model, and thus the distribution
of Lee-Yang zeros in this case are well-studied. As a matter of fact, all Lee-Yang zeros will be distributed on the unit circle in
this case as long as the coefficient of the nonlinear part is negative. However, the distribution behaviors of Lee-Yang zeros for
a higher nonlinearity are still unknown, even in the aspect of mathematics. To investigate it, in this paper we propose a nonlinear
toy model for quantum spins and thoroughly discuss the distribution of corresponding Lee-Yang zeros, especially whether they sit on
the unit circle.

Furthermore, similar to the previous studies on the detection of Lee-Yang zeros~\cite{Wei2012,Peng2015,Kuzmak2019,Wei2017}, we also
discuss the scenario that a probe qubit is coupled to the nonlinear system and show how to detect all Lee-Yang zeros by tuning
the coupling strength and the coefficient of the linear part in the nonlinear system. In the meantime, due to the fact that the density
matrix of this probe qubit is dependent on the temperature and coupling strength, the expression of the quantum Fisher information
matrix with respect to these two parameters at the Lee-Yang zeros is analytically calculated. Through the analysis of the quantum
Fisher information matrix, some interesting phenomena are discovered.

\section{The model and distribution of Lee-Yang zeros}

Consider the following nonlinear Hamiltonian
\begin{equation}
\gamma H^k_0+h H_0,
\end{equation}
where $\gamma$ and $h$ are constant coefficients for the nonlinear and linear parts. $k$ is the nonlinearity. Denote $\ket{n}$
as the eigenstate of $H_0$ with the eigenvalue $n$. In the case that the Hamiltonian is non-degenerate, the partition function
$Z=\mathrm{Tr}(e^{-\beta H})$ of this Hamiltonian can be written in a polynomial form,
\begin{equation}
Z=\sum_n p_n z^n,
\end{equation}
where $z=e^{-\beta h}$ and $p_n=e^{-\beta\gamma n^k}$. Here $\beta=1/(k_{\mathrm{B}}T)$ with $k_{\mathrm{B}}$ the Boltzmann
constant and $T$ the temperature. In the case that the Hamiltonian is degenerate, i.e., there exist $d_n$ eigenstates
$\ket{n_1},\ket{n_2},\dots,\ket{n_{d_n}}$ with respect to the eigenvalue $n$, then $p_n$ becomes $p_n=d_n e^{-\beta \gamma n^k}$.
Now consider a specific Hamiltonian form
\begin{equation}
H=\gamma J_z^k +h J_z.
\label{eq:H_Jz}
\end{equation}
Here $J_z=\frac{1}{2}\sum_{i=1}^{N}\sigma^z_j$ is the collective spin operator with $\sigma^z_j=\ket{\uparrow}\bra{\uparrow}-
\ket{\downarrow}\bra{\downarrow}$ the Pauli Z matrix for the $j$th spin. The state $\ket{\uparrow}$ ($\ket{\downarrow}$)
represents the spin up (down) state and $N$ is the number of spins. This Hamiltonian could be treated as the
generalized nonlinear collective spin system, and when $k=2$, it is nothing but the generalized one-axis twisting
model~\cite{Kitagawa1993,Ma2011,Jin2009}. The physical realization of this toy model for $k\geq3$ is still an open
question for now and requires further investigation. It is easy to see that the state $\bigotimes^N_{j=1}\ket{a_j}$
($a_j=\uparrow,\downarrow$) is the eigenstate of $J_z$ with respect to the eigenvalue $\frac{1}{2}(n_{\uparrow}-n_{\downarrow})$
with $n_{\uparrow}$ ($n_{\downarrow}$) the number of spin-up (-down) states in $\bigotimes^N_{j=1}\ket{a_j}$. As a matter of
fact, another well-known representation of the eigenstate of $J_z$ is the Dicke state $\ket{J,m}$ and the corresponding
eigenvalue is $m$. Here $J$ is the total angular momentum. Further defining $n:=m+J$ ($n=0,1,\dots,2J$), the Dicke state can
be rewritten into $\ket{n}:=\ket{J,n-J}$, and the degeneracy of $\ket{n}$ is $\binom{N}{n}=\frac{N!}{n!(N-n)!}$, the binomial
coefficient. Utilizing the basis $\{\ket{n}\}$, the partition function for the Hamiltonian (\ref{eq:H_Jz}) can be expressed by
\begin{equation}
Z=e^{\frac{1}{2}\beta h N}\sum_{n=0}^{N}\binom{N}{n}e^{-\beta\gamma\left(n-\frac{N}{2}\right)^k} z^n
\label{eq:pf}
\end{equation}
with $z:=e^{-\beta h}$. Hence the partition function can be viewed as an $N$th order polynomial function of $z$. Utilizing
the roots $\{z_i\}^{N}_{i=1}$ of the equation $Z(z)=0$, the expression above can be factorized to
\begin{eqnarray}
Z=e^{\frac{1}{2}\beta h N-\beta \gamma\left(-\frac{N}{2}\right)^k}\prod_{i=1}^{N}\left(z-z_{i}\right).
\label{eq:pf1}
\end{eqnarray}

\begin{figure*}[tp]
\centering
\includegraphics[width=18cm]{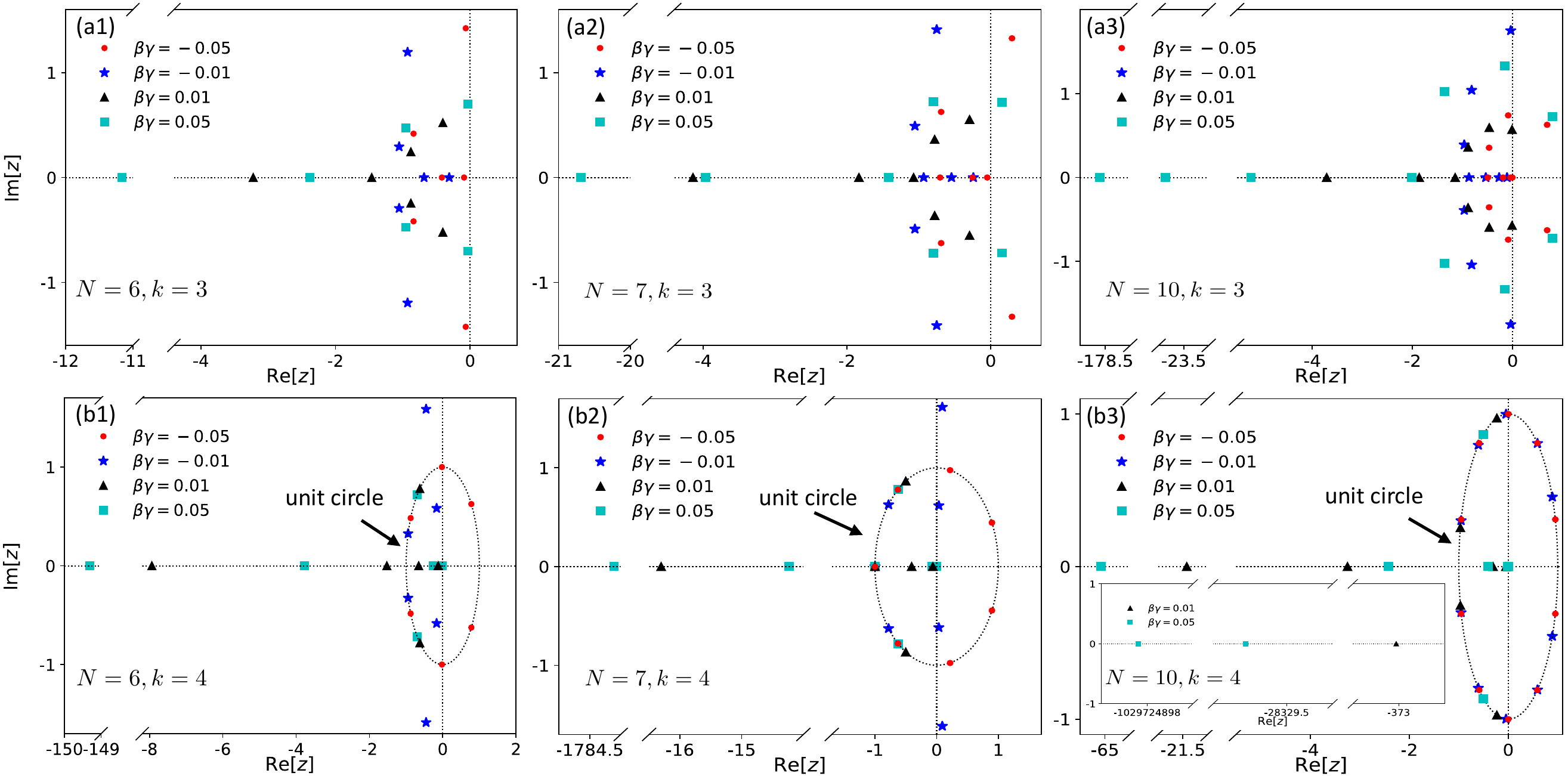}
\caption{Distribution of Lee-Yang zeros for the nonlinearity $k=3$ in the case of (a1) $N=6$, (a2) $N=7$,  and (a3) $N=10$ 
and for the nonlinearity $k=4$ in the case of (b1) $N=6$, (b2) $N=7$, and (b3) $N=10$. The red circles, blue pentagrams, 
black triangles, and cyan squares represent the Lee-Yang zeros for $\beta\gamma=-0.05$, $-0.01$, $0.01$, and $0.05$, 
respectively. The inset of (b3) shows the Lee-Yang zeros that are not presented in (b3).}
\label{Fig:LY}
\end{figure*}

A more interesting fact is that $z$ can be extended to the complex plane via the analytic continuation, which means the
solutions of $Z(z)=0$ are also extended to the complex plane. These roots on the complex plane are usually referred to as
the Lee-Yang zeros. Equation~(\ref{eq:pf1}) indicates that the property of the partition function can be reflected by the
roots $\{z_i\}^{N}_{i=1}$. Now let us study the behaviors of the distribution of $\{z_i\}^{N}_{i=1}$. One can see from
Eq.~(\ref{eq:pf}) that $e^{\frac{1}{2}\beta h N}$ is a global coefficient and does not affect the solutions of $Z(z)=0$,
indicating that the distribution of $\{z_i\}^{N}_{i=1}$ is independent of $\beta h$. The distributions of Lee-Yang zeros for
different values of $\beta\gamma$ for the nonlinearity $k=3$ ($k=4$) in the case of $N=6, 7$, and $10$ are illustrated 
in Figs.~\ref{Fig:LY}(a1)-\ref{Fig:LY}(a3) [Figs.~\ref{Fig:LY}(b1)-\ref{Fig:LY}(b3)].

In all cases, the distributions of Lee-Yang zeros for all values of $\beta\gamma$, including $\beta\gamma=-0.05$
(red circles), $\beta\gamma=-0.01$ (blue pentagrams), $\beta\gamma=0.01$ (black triangles), and $\beta\gamma=0.05$ (cyan
squares), are all symmetric about the axis of $\mathrm{Re}[z]$. Here $\mathrm{Re}[\cdot]$ and $\mathrm{Im}[\cdot]$ represent
the real and imaginary part. A more interesting phenomenon is that the point $(-1,0)$ is always a Lee-Yang zero in the case
of $k=4$. As a matter of fact, this result can be generalized to the case with an odd $N$ and even nonlinearity $k$, as given
in the theorem below.

\begin{figure}[tp]
\centering
\includegraphics[width=8.6cm]{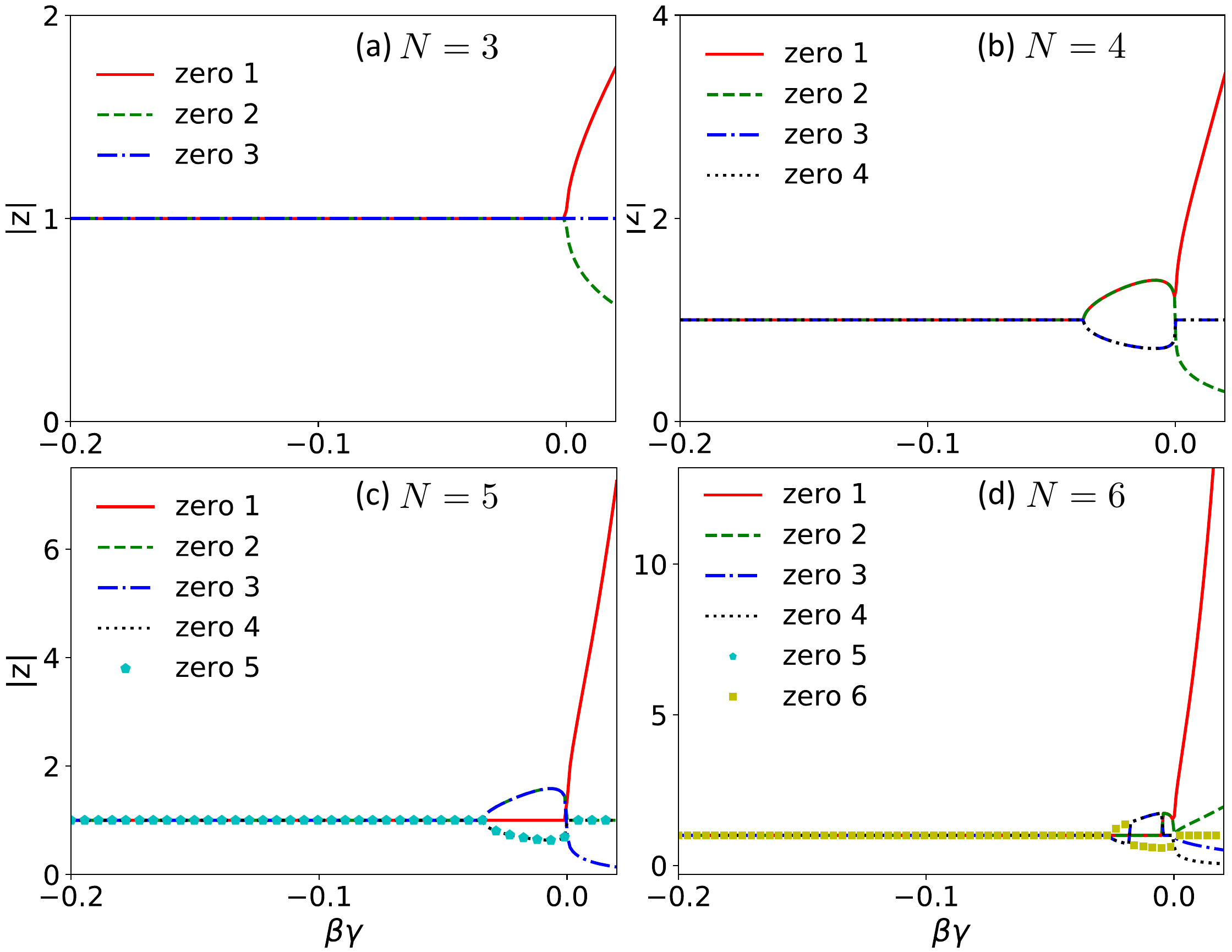}
\caption{Norms of all Lee-Yang zeros in the case of $k=4$ for different spin numbers. The spin numbers 
are (a) $N=3$, (b) $N=4$, (c) $N=5$, and (d) $N=6$. Zero 1 to zero 6 in the plots are the labels of the Lee-Yang zeros.}
\label{Fig:unit}
\end{figure}

\emph{Theorem 1.} For the Hamiltonian (\ref{eq:H_Jz}), the point $(-1,0)$ in the complex plane is always a
Lee-Yang zero when the spin number $N$ is odd and the nonlinearity $k$ is even.

This theorem can be proved by noticing that
\begin{align*}
& \sum_{n=0}^{N}\binom{N}{n}e^{-\beta\gamma\left(n-\frac{N}{2}\right)^k} (-1)^n \\
=& \sum_{n=0}^{\frac{N-1}{2}}\binom{N}{n}\!\!\left[e^{-\beta\gamma\left(n-\frac{N}{2}\right)^k} (-1)^n
\!+\!e^{-\beta\gamma\left(\frac{N}{2}-n\right)^k} (-1)^{N-n}\right]\!,
\end{align*}
where the equality $\binom{N}{n}=\binom{N}{N-n}$ was applied. In the case that $k$ is even, the equation above further
reduces to
\begin{equation}
\sum_{n=0}^{\frac{N-1}{2}}\binom{N}{n}e^{-\beta\gamma\left(n-\frac{N}{2}\right)^k}
\left[(-1)^n+(-1)^{N-n}\right]
\end{equation}
When $N$ is odd, $(-1)^n+(-1)^{N-n}$ is always zero. The theorem is then proved. $\hfill\blacksquare$

In the case of $k=2$, all Lee-Yang zeros will be on the unit circle as long as $\beta\gamma$ is negative~\cite{Wei2012,Peng2015}.
However, as shown in Fig.~\ref{Fig:LY}, the situation becomes complex when $k$ is larger than $2$. In the case that
$k$ is odd, we have the following theorem.

\emph{Theorem 2.} For the Hamiltonian (\ref{eq:H_Jz}), the Lee-Yang zeros are never all distributed on the unit
circle when the nonlinearity $k$ is odd.

According to Vieta's formulas, the Lee-Yang zeros $\{z_i\}$ satisfy 
\begin{equation}
\prod^N_{i=1}z_i=(-1)^N e^{\beta\gamma\left[\left(\frac{N}{2}\right)^k-\left(-\frac{N}{2}\right)^k\right]}.
\end{equation}
In the case that $k$ is odd, one can further have $\prod^N_{i=1}|z_i|=e^{2\beta\gamma\left(\frac{N}{2}\right)^k}$. It is
obvious that $e^{2\beta\gamma\left(\frac{N}{2}\right)^k}$ cannot be 1 as long as $\beta\gamma\neq 0$, indicating that
the Lee-Yang zeros cannot be all distributed on the unit circle when in this case. The theorem is then proved.
$\hfill\blacksquare$

From the proof above, one can immediately obtain the following theorem for an even nonlinearity.

\emph{Theorem 3.} For the Hamiltonian (\ref{eq:H_Jz}), the Lee-Yang zeros satisfy $\prod^N_{i=1}|z_i|=1$
when the nonlinearity $k$ is even.

This theorem does not lead to the result that all Lee-Yang zeros are distributed on the unit circle when $k$ is even,
which is already exhibited in Fig.~\ref{Fig:LY}(b). In the case of $k=4$, we find an interesting phenomenon for
$N=3,4,5,6$ that the norms of all Lee-Yang zeros are $1$, namely, all Lee-Yang zeros are distributed on the unit circle,
when $\beta\gamma$ is smaller than a critical value, as shown in Figs.~\ref{Fig:unit}(a) to \ref{Fig:unit}(d) for $N=3$,
$N=4$, $N=5$, and $N=6$, respectively. As a matter of fact, when $k=4$, the Lee-Yang zeros will always be distributed on
the unit circle as long as $\beta\gamma$ is small enough, regardless of the value of $N$. This is due to the fact that
when $k=4$, the equation $Z(z)=0$ reduces to
\begin{equation}
\sum^N_{n=0}\binom{N}{n}e^{\beta\gamma\left[\frac{N^4}{16}-\left(n-\frac{N}{2}\right)^4\right]}z^n=0.
\end{equation}
It is obvious that
\begin{equation}
\frac{N^4}{16}-\left(n-\frac{N}{2}\right)^4=\frac{N^4}{16}\left[1-\left(\frac{2n}{N}-1\right)^4\right]\geq 0
\end{equation}
for $n\in [0,N]$ since $n/N\leq 1$. Hence, when $\beta\gamma$ is small enough, namely, $\beta\gamma$ is negative and
its absolute value is large enough, $e^{\beta\gamma[\frac{N^4}{16}-(n-\frac{N}{2})^4]}\approx 0$ and the equation above
approximates to
\begin{equation}
1+z^N=0,
\end{equation}
which immediately gives $|z|=1$, indicating that the Lee-Yang zeros are distributed on the unit circle. As a matter of fact,
this result can be extended to the case of all even values of nonlinearity. In this case, the equation $Z(z)=0$ reduces to
\begin{equation}
\sum^N_{n=0}\binom{N}{n}e^{\beta\gamma\left[\left(\frac{N}{2}\right)^k-\left(n-\frac{N}{2}\right)^k\right]}z^n=0,
\end{equation}
where $\left(\frac{N}{2}\right)^k-\left(n-\frac{N}{2}\right)^k\geq 0$. Therefore, when $\beta\gamma$ is small enough,
the equation above always approximates to $1+z^N=0$ and the Lee-Yang zeros are thus distributed on
the unit circle. Hence, we have the following theorem.

\emph{Theorem 4.} For the Hamiltonian (\ref{eq:H_Jz}), the Lee-Yang zeros are always distributed on the unit circle for an
even nonlinearity $k$ as long as $\beta\gamma$ is small enough.

In Figs.~\ref{Fig:unit}(a) to \ref{Fig:unit}(d), there exists a critical value of $\beta\gamma$ for all zeros to be simultaneously
distributed on the unit circle. Whether this critical point exists in general and how to analytically obtain this critical point
are not answered in the theorem above and still remain open questions that require further investigations.

\section{Detection of Lee-Yang zeros with single qubit}

\begin{figure*}[tp]
\centering\includegraphics[width=18cm]{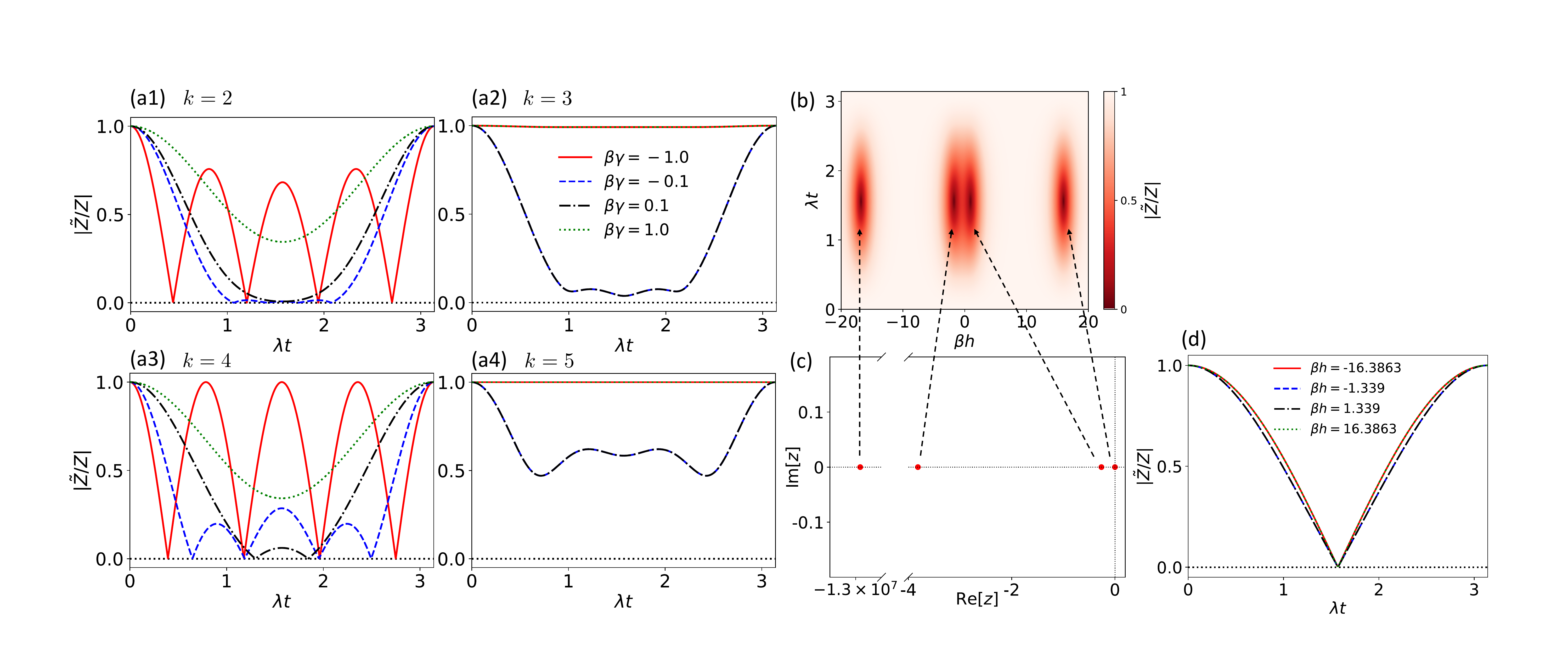}
\caption{Detection of Lee-Yang zeros in the case of $N=4$. (a1)-(a4) The evolution of $|\tilde{Z}/Z|$ for (a1) $k=2$, 
(a2) $k=3$, (a3) $k=4$, and (a4) $k=5$. The solid red, dashed blue, dash-dotted black, and dotted green lines represent 
the values of amplitudes for $\beta\gamma=-1.0$, $-0.1$, $0.1$, and $1.0$, respectively. $\beta h$ is set to be zero in
(a1)-(a4).  (b) The values of $|\tilde{Z}/Z|$ as a function of $\beta h$ and $\lambda t$. (c) The distribution of Lee-Yang 
zeros. (d) The evolution of $|\tilde{Z}/Z|$ for the values of $\beta h$ to reach the Lee-Yang zeros. In (b)-(d) the nonlinearity 
$k=4$ and $\beta\gamma=1.0$. }
\label{fig:qubit}
\end{figure*}

\begin{figure*}[tp]
\centering\includegraphics[width=18cm]{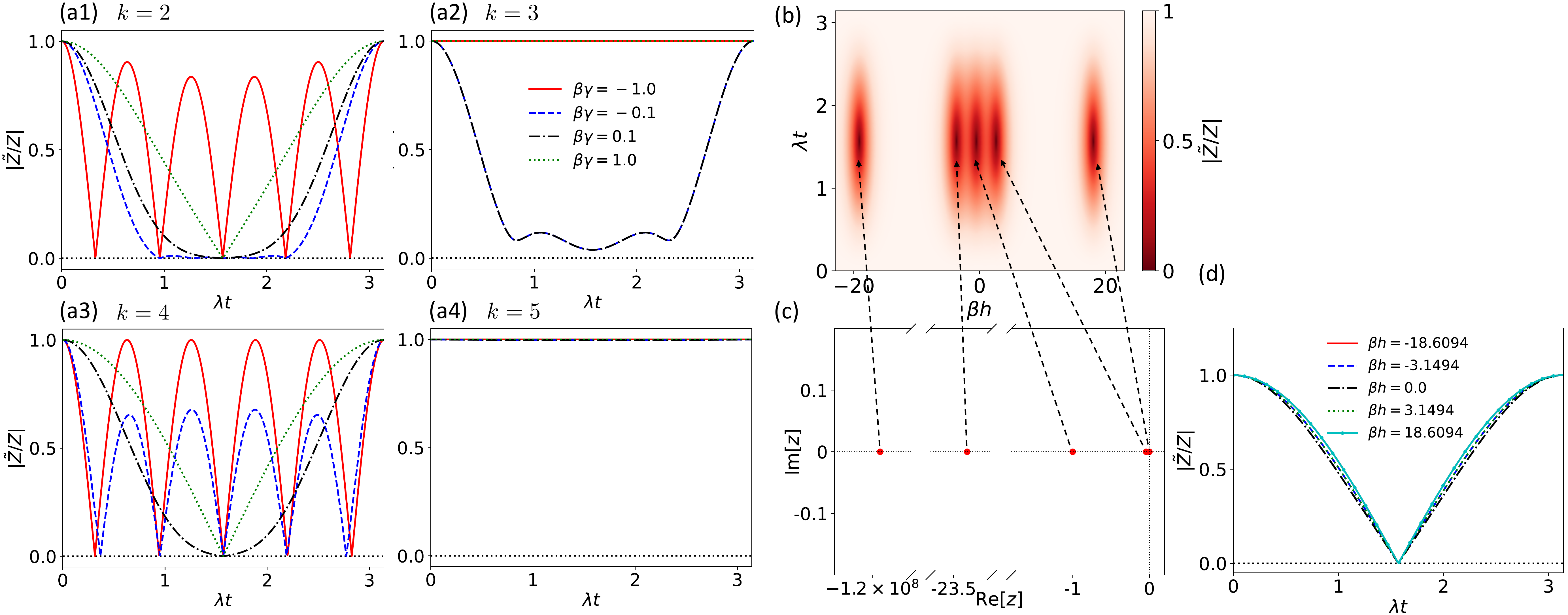}
\caption{Detection of Lee-Yang zeros in the case of $N=5$.  (a1)-(a4) The evolution of $|\tilde{Z}/Z|$ for (a1) $k=2$,  
(a2) $k=3$, (a3) $k=4$, and (a4) $k=5$. The solid red, dashed blue, dash-dotted black, and dotted green lines represent 
the values of amplitudes for $\beta\gamma=-1.0$, $-0.1$, $0.1$, and $1.0$, respectively. $\beta h$ is set to be zero in
(a1)-(a4).  (b) The values of $|\tilde{Z}/Z|$ as a function of $\beta h$ and $\lambda t$. (c) The distribution of Lee-Yang 
zeros. (d) The evolution of $|\tilde{Z}/Z|$ for the values of $\beta h$ to reach the Lee-Yang zeros. In (b)-(d) the 
nonlinearity $k=4$ and $\beta\gamma=0.5$. }
\label{fig:qubit_N5}
\end{figure*}

The scheme of detecting Lee-Yang zeros with a probe qubit is first proposed by Wei and Liu in 2012~\cite{Wei2012} and
was further simulated in experiments by Peng \emph{et al.} in 2015~\cite{Peng2015}. Here we also consider the coupling
between a probe state and Hamiltonian (\ref{eq:H_Jz}) and discuss the detection of Lee-Yang zeros. The total Hamiltonian is
\begin{equation}
H_{\mathrm{tot}}=H+\frac{1}{2}\omega_0\sigma_z+\lambda J_z\sigma_z,
\end{equation}
where $H$ is given in Eq.~(\ref{eq:H_Jz}), $\omega_0$ is the frequency of the probe qubit, and $\lambda$ is the
coupling strength between it and the nonlinear system. Now denote the total Hilbert space as
$\mathcal{H}_{\mathrm{tot}}=\mathcal{H}_{\mathrm{q}}\otimes\mathcal{H}$ with $\mathcal{H}_{\mathrm{q}}$ and
$\mathcal{H}$ the Hilbert space of the probe qubit and nonlinear system. In this way, $J_z$ here actually represents
$\openone_{\mathrm{q}}\otimes J_z$, and $\sigma_z$ represents $(\ket{\uparrow}\bra{\uparrow}-\ket{\downarrow}\bra{\downarrow})
\otimes\openone$ with $\openone_{\mathrm{q}}$ and $\openone$ the identity operators in $\mathcal{H}_{\mathrm{q}}$
and $\mathcal{H}$. Assume the initial state is a product state
\begin{equation}
\rho_{\mathrm{in}}=\rho_0\otimes\rho_{\mathrm{th}},
\end{equation}
where $\rho_0$ is the initial state of the probe qubit and $\rho_{\mathrm{th}}=e^{-\beta H}/Z$ is the thermal state of the
nonlinear system.

The evolved state $\rho_t$ for the probe qubit can be calculated via the equation
\begin{equation}
\rho_t=\mathrm{Tr}_{\mathcal{H}}\left(e^{-i H_{\mathrm{tot}}t}\rho_{\mathrm{in}}e^{i H_{\mathrm{tot}}t}\right),
\end{equation}
where $\mathrm{Tr}_{\mathcal{H}}(\cdot)$ represents the partial trace on the nonlinear system. Utilizing this equation and
realizing that
\begin{equation}
e^{-i\lambda t J_z \sigma_z}=\cos(\lambda t J_z)\openone_{\mathrm{tot}}-i\sin(\lambda t J_z)\sigma_z
\end{equation}
with $\openone_{\mathrm{tot}}$ the identity operator in $\mathcal{H}_{\mathrm{tot}}$, $\rho_t$ can be solved analytically.
In the basis $\{\ket{\uparrow},\ket{\downarrow}\}$, $\rho_t$ can be expressed by
\begin{equation}
\rho_t=\left(\begin{array}{cc}
[\rho_0]_{00} & \frac{\tilde{Z}}{Z}e^{-i\omega_0 t}[\rho_0]_{01} \\
\frac{\tilde{Z}^{*}}{Z}e^{i\omega_0 t}[\rho_0]_{10} & [\rho_0]_{11} \\
\end{array}\right),
\label{eq:rhot}
\end{equation}
where $[\rho_0]_{ij}$ is the $ij$th entry of $\rho_0$ and
\begin{equation}
\tilde{Z}=\mathrm{Tr}\left(e^{-\beta H-i2\lambda tJ_z}\right).
\end{equation}
Similar to the partition function $Z$, $\tilde{Z}$ can also be expressed by
\begin{equation}
\tilde{Z}=e^{\left(\frac{1}{2}\beta h+i\lambda t\right)N}\sum_{n=0}^{N}\binom{N}{n}
e^{-\beta\gamma\left(n-\frac{N}{2}\right)^k}\tilde{z}^{n}
\end{equation}
with $\tilde{z}:=e^{-\beta h-i2\lambda t}$. Compared to Eq.~(\ref{eq:pf}), it is not difficult to see that the equations
$Z(z)=0$ and $\tilde{Z}(\tilde{z})=0$ share the same solutions. Hence, $\tilde{Z}$ can also be factorized to
\begin{equation}
\tilde{Z}=e^{\left(\frac{1}{2}\beta h+i\lambda t\right)N-\beta\gamma\left(-\frac{N}{2}\right)^k}
\prod_{i=1}^{N}\left(\tilde{z}-z_i\right).
\label{eq:pf_bit1}
\end{equation}

As shown in Eq.~(\ref{eq:rhot}), the nonlinear system is responsible for the evolution of the nondiagonal entries of $\rho_t$,
indicating that the information of the Lee-Yang zeros $\{z_i\}$ is hidden in the dynamics of the probe qubit. Utilizing
Eqs.~(\ref{eq:pf1}) and (\ref{eq:pf_bit1}), the amplitude $|\tilde{Z}e^{-i\omega_0 t}/Z|=|\tilde{Z}/Z|$ reduces to
\begin{equation}
\left|\frac{\tilde{Z}}{Z}\right|=\left|\frac{\prod_{i=1}^{N}\left(\tilde{z}-z_{i}\right)}
{\prod_{i=1}^{N}\left(z-z_i\right)}\right|.
\end{equation}
From this expression, it can be seen that this amplitude above vanishes when $\tilde{z}$ reaches the zeros $\{z_i\}$. Hence, the
zeros can be measured via the evolution of $|\tilde{Z}/Z|$ as long as it can vanish. The evolution of $|\tilde{Z}/Z|$ for different
nonlinearity in the case of $\beta h=0$ is given in Figs.~\ref{fig:qubit}(a1) to \ref{fig:qubit}(a4) [Figs.~\ref{fig:qubit_N5}(a1) to 
\ref{fig:qubit_N5}(a4)] for $N=4$ ($N=5$). It can be seen that the Lee-Yang zeros can be easily detected via the amplitude 
$|\tilde{Z}/Z|$ when the nonlinearity $k$ is even, as shown in Figs.~\ref{fig:qubit}(a1) and \ref{fig:qubit_N5}(a1) for $k=2$ and 
Figs.~\ref{fig:qubit}(a3) and \ref{fig:qubit_N5}(a3) for $k=4$, especially when $\beta\gamma$ is negative. With the increase of 
$\beta\gamma$, it gets difficult for $|\tilde{Z}/Z|$ to vanish, indicating that the Lee-Yang zeros cannot be detected via the 
amplitude $|\tilde{Z}/Z|$ in this parameter region. In the case that $k$ is odd, as illustrated in Figs.~\ref{fig:qubit}(a2) and 
\ref{fig:qubit_N5}(a2) for $k=3$ and Figs.~\ref{fig:qubit}(a4) and \ref{fig:qubit_N5}(a4) for $k=5$, $|\tilde{Z}/Z|$ can hardly 
vanish, especially when the norm of $\beta\gamma$ is large. These phenomena indicate that some value regions of $\beta\gamma$ 
could be unfriendly for the detection of Lee-Yang zeros in this case. Then how to detect the Lee-Yang zeros in these regions of 
$\beta\gamma$ becomes a serious problem. Luckily, the distribution of Lee-Yang zeros $\{z_i\}^N_{i=1}$ does not rely on the values 
of $h$, yet the amplitude $|\tilde{Z}/Z|$ is dependent on it, which provides a method to further detect the Lee-Yang zeros in these cases.

We demonstrate this detection strategy for the nonlinearity $k=4$ in both cases of $N=4, \beta\gamma=1.0$ and $N=5, \beta\gamma=0.5$, 
as given in Figs.~\ref{fig:qubit}(b) to \ref{fig:qubit}(d) and Figs.~\ref{fig:qubit_N5}(b) to \ref{fig:qubit_N5}(d). From Fig.~\ref{fig:qubit}(b) 
[Fig.~\ref{fig:qubit_N5}(b)], it can be seen that four vanishing points of $|\tilde{Z}/Z|$ are shown at the time $\lambda t=\pi/2$ when the 
values of $\beta h$ are changed from around $-20$ to $20$. These vanishing points correspond to the four Lee-Yang zeros in this case, 
as shown in Fig.~\ref{fig:qubit}(c) [Fig.~\ref{fig:qubit_N5}(c)]. The reason why the zeros always occur at the time $\pi/2$ is due to the fact 
that all four Lee-Yang zeros are located on the negative axis of $\mathrm{Re}[z]$. To make sure $e^{-\beta h-i2\lambda t}$ is real and 
negative, the only available value of $\lambda t$ is $\lambda t=\pi/2$. In the meantime, in this case the proper values of $\beta h$ for the 
detection of Lee-Yang zeros are $\{-\ln|z_i|\}$, and the evolution of $|\tilde{Z}/Z|$ with $\beta h\in \{-\ln|z_i|\}$ are shown in 
Fig.~\ref{fig:qubit}(d) [Fig.~\ref{fig:qubit_N5}(d)]. The vanishing points indeed always occur at the time $\pi/2$ and the Lee-Yang zeros 
are then detectable.

\section{Quantum Fisher information matrix at the Lee-Yang zeros}

Quantum Fisher information matrix is another important fundamental quantity in quantum mechanics and quantum information. In this
section we discuss the behaviors of quantum Fisher information matrix of the probe qubit at the Lee-Yang zeros. For the evolved
state in Eq.~(\ref{eq:rhot}), the quantum Fisher information matrix for the parameters $\{\lambda,\beta\}$ can be calculated
via the equation~\cite{Dittmann1999,Liu2020}
\begin{equation}
\mathcal{F}_{ab}=2\mathrm{Tr}\left[(\partial_a\rho_t)(\partial_b\rho_t)\right]
\label{eq:QFIM_pure}
\end{equation}
for a pure $\rho_t$, and
\begin{align}
\mathcal{F}_{ab} =&~ \mathrm{Tr}\left[\left(\partial_a \rho_t\right)\left(\partial_b \rho_t\right)\right] \nonumber \\
& +\frac{1}{\det(\rho_t)}\mathrm{Tr}\left[\rho_t\left(\partial_a \rho_t\right)\rho_t\left(\partial_b \rho_t\right)\right]
\label{eq:QFIM_mix}
\end{align}
for a mixed $\rho_t$. The subscripts $a,b\in\{\lambda,\beta\}$. Next, denoting $g=\tilde{Z}/Z$, it can be seen that
\begin{equation}
\partial_{\lambda(\beta)}g=g\Delta E_{\lambda(\beta)}:=g(E_{\lambda(\beta)}-\tilde{E}_{\lambda(\beta)})
\label{eq:pg}
\end{equation}
with $E_{\lambda(\beta)}=-\partial_{\lambda(\beta)}\ln Z$ and $\tilde{E}_{\lambda(\beta)}=-\partial_{\lambda(\beta)}\ln\tilde{Z}$.
Here $E_{\beta}$ is nothing but the thermodynamic energy for the Hamiltonian (\ref{eq:H_Jz}). $E_{\lambda}=0$ due to the fact
that $Z$ is independent of $\lambda$. Utilizing Eqs.~(\ref{eq:QFIM_pure}) and (\ref{eq:pg}), the entries of the quantum Fisher
information matrix are of the form
\begin{align}
\mathcal{F}_{\lambda\lambda(\beta\beta)}=&~4|g|^2|[\rho_0]_{01}|^2|\Delta E_{\lambda(\beta)}|^2,
\label{eq:QFIM_pure1} \\
\mathcal{F}_{\lambda\beta} =&~ 4|g|^2|[\rho_0]_{01}|^2\mathrm{Re}\left[\Delta E_{\lambda}(\Delta E_{\beta})^*\right],
\label{eq:QFIM_pure2}
\end{align}
when $\rho_t$ is pure. And when $\rho_t$ is mixed, they are
\begin{align}
\mathcal{F}_{\lambda\lambda(\beta\beta)} =&~ 4|g|^2|[\rho_0]_{01}|^2\Bigg(\left|\Delta E_{\lambda(\beta)}\right|^2 \nonumber \\
& +\frac{|g|^2|[\rho_0]_{01}|^2\mathrm{Re}^2\left[\Delta E_{\lambda(\beta)}\right]}
{[\rho_0]_{00}[\rho_0]_{11}-|g|^2|[\rho_0]_{01}|^2}\Bigg),
\label{eq:QFIM_mix1} \\
\mathcal{F}_{\lambda\beta} =&~ 4|g|^2|[\rho_0]_{01}|^2\Bigg(\mathrm{Re}\left[\Delta E_{\lambda}(\Delta E_{\beta})^{*}\right]
\nonumber \\
& +\frac{|g|^2|[\rho_0]_{01}|^2\mathrm{Re}\left[\Delta E_{\lambda}\right]
\mathrm{Re}\left[\Delta E_{\beta}\right]}{[\rho_0]_{00}[\rho_0]_{11}-|g|^2|[\rho_0]_{01}|^2}\Bigg).
\label{eq:QFIM_mix2}
\end{align}

For the sake of investigating the general behaviors of the quantum Fisher information matrix at the Lee-Yang zeros, its general
expression at these points should be provided. As a matter of fact, the value of $\tilde{Z}$ is zero when the zero of $g$ reaches
a Lee-Yang zero. Taking $(\ket{\uparrow}+\ket{\downarrow})/\sqrt{2}$ as the initial state of the probe qubit and utilizing the
condition $\tilde{Z}=0$, the entries of the quantum Fisher information matrix for both pure and mixed $\rho_t$ at the Lee-Yang
zeros can be written as
\begin{eqnarray}
\mathcal{F}_{\lambda\lambda(\beta\beta)} &=& \frac{1}{Z^2}\left|\partial_{\lambda(\beta)}\tilde{Z}\right|^2, \\
\mathcal{F}_{\lambda\beta} &=& \frac{1}{Z^2}\mathrm{Re}\left[(\partial_{\lambda}\tilde{Z})(\partial_{\beta}\tilde{Z}^*)\right].
\end{eqnarray}
Notice that the zero of $g$ can only reach one Lee-Yang zero with a group of specific values of $h$ and $\lambda$, which allows
us to assume, without loss of generality, that Lee-Yang zero is $m$th zero ($z_m$), namely, $e^{-\beta h}\cos(2\lambda t)=\mathrm{Re}[z_m]$
and $e^{-\beta h}\sin(2\lambda t)=\mathrm{Im}[z_m]$. In the following we denote $h_m$ and $\lambda_m$ as the values of $h$ and
$\lambda$ that satisfy these equations. In this case, $\partial_{\lambda}\tilde{Z}$ can be expressed by
\begin{eqnarray*}
\partial_{\lambda}\tilde{Z} &=& -i2te^{(\frac{1}{2}\beta h_m+i\lambda_m t)N-\beta\gamma\left(-\frac{N}{2}\right)^k}
\prod_{i\neq m}(\tilde{z}-z_i)z_m, \\
\partial_{\beta}\tilde{Z} &=& -h_m e^{(\frac{1}{2}\beta h_m+i\lambda_m t)N-\beta\gamma\left(-\frac{N}{2}\right)^k}
\prod_{i\neq m}(\tilde{z}-z_i)z_m.
\end{eqnarray*}
Hence, the entries of the quantum Fisher information matrix can be rewritten into
\begin{eqnarray}
\mathcal{F}_{\lambda\lambda} &=& 4t^2e^{-2\beta h_m}\frac{\prod_{i\neq m}|(z_m-z_i)|^2}
{\prod^N_{i=1}(|z_m|-z_i)^2}, \\
\mathcal{F}_{\beta\beta} &=& h_m^2 e^{-2\beta h_m} \frac{\prod_{i\neq m}|(z_m-z_i)|^2}
{\prod^N_{i=1}(|z_m|-z_i)^2}, \\
\mathcal{F}_{\lambda\beta} &=& 0.
\end{eqnarray}
From the perspective of quantum parameter estimation, $\mathcal{F}_{\lambda\beta}=0$ means that in theory, the optimal measurement
can let the deviations of $\lambda$ and $\beta$ reach their precision limit simultaneously. Furthermore, when $z_m$ sits on the unit
circle, $h$ has to be $0$ and $\mathcal{F}_{\beta\beta}$ vanishes. This result indicates that the probe qubit cannot work as the
thermometer at the position of a Lee-Yang zero if this zero is on the unit circle.

Next, let us discuss a more specific regime that $\beta\gamma$ is small. Theorem 4 shows that in this regime the Lee-Yang zeros are
always distributed on the unit circle for even nonlinearity. In this case, $Z$ and $\tilde{Z}$ can be approximated into
\begin{align}
Z &\approx 2e^{-\beta\gamma\left(\frac{N}{2}\right)^k}\cosh(\frac{1}{2}\beta hN), \label{eq:Zapprox} \\
\tilde{Z} &\approx 2e^{-\beta\gamma\left(\frac{N}{2}\right)^k}\cosh(\frac{1}{2}\beta hN+i\lambda t N). \label{eq:Ztapprox}
\end{align}
Utilizing Eqs.~(\ref{eq:Zapprox}) and (\ref{eq:Ztapprox}), $|g|^2$ can be written as
\begin{equation}
|g|^2=1-\frac{\sin^2(\lambda tN)}{\cosh^2\left(\frac{1}{2}\beta hN\right)}.
\end{equation}
In the meantime, $\tilde{E}_{\lambda}$ and $\tilde{E}_{\beta}$ read
\begin{eqnarray}
\tilde{E}_{\lambda} &=& tN\frac{\sin(2\lambda tN)-i\sinh(\beta hN)}{\cosh(\beta hN)+\cos(2\lambda tN)}, \\
\tilde{E}_{\beta} &=& \!\gamma\left(\frac{N}{2}\right)^k\!\!-\frac{1}{2}hN\frac{\sinh(\beta hN)+i\sin(2\lambda tN)}
{\cosh(\beta hN)+\cos(2\lambda tN)}.
\end{eqnarray}
Due to the fact that $E_{\lambda}=0$ and
\begin{equation}
E_{\beta}=\gamma\left(\frac{N}{2}\right)^k-\frac{1}{2}hN\tanh\left(\frac{1}{2}\beta h N\right),
\end{equation}
one can immediately have
\begin{eqnarray*}
\Delta E_{\lambda} &=& -tN\frac{\sin(2\lambda tN)-i\sinh(\beta hN)}{\cosh(\beta hN)+\cos(2\lambda tN)}, \\
\Delta E_{\beta} &=& \frac{1}{2}hN\frac{2\sin^2(\lambda tN)\tanh\left(\frac{1}{2}\beta h N\right)+i\sin(2\lambda tN)}
{\cosh(\beta hN)+\cos(2\lambda tN)}.
\end{eqnarray*}
Still taking the initial state of the probe qubit as $(\ket{\uparrow}+\ket{\downarrow})/\sqrt{2}$, the entries of the quantum Fisher
information matrix for a pure $\rho_t$ [Eqs.~(\ref{eq:QFIM_pure1}) and (\ref{eq:QFIM_pure2})] can be expressed by
\begin{eqnarray}
\mathcal{F}_{\lambda\lambda} &=& t^2 N^2 \left[1-\frac{\cos^2(\lambda tN)}{\cosh^2(\frac{1}{2}\beta hN)}\right], \\
\mathcal{F}_{\beta\beta} &=& \frac{1}{4}h^2 N^2\frac{\sin^2(\lambda tN)}{\cosh^4(\frac{1}{2}\beta hN)}, \\
\mathcal{F}_{\lambda\beta} &=& \frac{1}{2}htN^2\frac{\sin(2\lambda tN)\sinh(\beta hN)}{[1+\cosh(\beta hN)]^2}.
\end{eqnarray}
It is obvious that $\rho_t$ can only be pure when $|g|^2=1$, i.e., $\sin(\lambda tN)=0$, which means the expression of the quantum Fisher
information matrix for pure states is only valid for some specific time points. And these points may not correspond to the Lee-Yang
zeros. Hence, in the following we only discuss the case that $\rho_t$ is mixed. For a mixed $\rho_t$, the entries [Eqs.~(\ref{eq:QFIM_mix1})
and (\ref{eq:QFIM_mix2})] read
\begin{eqnarray}
\mathcal{F}_{\lambda\lambda} &=& t^2 N^2,   \\
\mathcal{F}_{\beta\beta} &=& \frac{1}{4}h^2 N^2\frac{\sin^2(\lambda tN)}
{\cosh^2\left(\frac{1}{2}\beta hN\right)}, \\
\mathcal{F}_{\lambda\beta} &=& 0.
\end{eqnarray}
The fact that both $\mathcal{F}_{\lambda\lambda}$ and $\mathcal{F}_{\beta\beta}$ are proportional to $N^2$ indicates that although 
the evolved state is mixed, both the deviations of $\lambda$ and $\beta$ can beat the standard quantum limit, $1/\sqrt{N}$ in this case, 
and reach the scale of $1/N$. Standard quantum limit is an important precision limit and error scaling in quantum metrology. It usually 
represents the measurement capability of a classical apparatus, and beating it indicates that the estimations of $\lambda$ and $\beta$ 
with this nonlinear system would overperform, at least theoretically, many classical measurement apparatuses. 

Different from the behaviors of $g$, the dynamics of $\mathcal{F}_{\lambda\lambda}$ does not show any relevance with the
Lee-Yang zeros since it does not rely on the values of $\lambda$ and $h$. When the zero of $g$ reaches a Lee-Yang zero, the
value of $\mathcal{F}_{\lambda\lambda}$ has no difference from other points. With respect to $\mathcal{F}_{\beta\beta}$,
the phenomenon is the same as the aforementioned general discussion. In this case, the probe qubit cannot work as a thermometer
at any Lee-Yang zero since all zeros are distributed on the unit circle, as stated in Theorem 4.

Although the probe qubit cannot be a thermometer at the Lee-Yang zeros, the direction of the zeros may still benefit the estimation
of $\beta$. For example, Theorem 1 tells us that the point $(-1,0)$ is always a Lee-Yang zero in this case as long as $N$ is odd.
On the direction of $(-1,0)$, the value of $\lambda t$ is $\pi/2+m\pi$ with $m$ a natural number. It is obvious that for these
values $\sin^2(\lambda tN)$ is $1$ since $N$ is odd, and $\mathcal{F}_{\beta\beta}$ reach its maximum value with respect to the time.

\section{Conclusion}

In summary, in this paper we proposed a nonlinear quantum spin model and discussed the distribution of the Lee-Yang zeros in this model.
Four observations are provided. For an odd nonlinearity, not all the Lee-Yang zeros can be distributed on the unit circle simultaneously.
In the case of an even nonlinearity, the point $(-1,0)$ is always a Lee-Yang zero when the spin number is odd. In the meantime, the
production of the norms of all Lee-Yang zeros is always 1, and when $\beta\gamma$ is small enough, all Lee-Yang zeros will always 
be distributed on the unit circle. Furthermore, the detection of these Lee-Yang zeros via a probe qubit is thoroughly discussed. In the case 
that the amplitude $|\tilde{Z}/Z|$ has no zero point during the dynamics, a detection scheme has been proposed via tuning the parameters 
$h$ and $\lambda$. Moreover, the quantum Fisher information matrix for $\lambda$ and $\beta$ at the Lee-Yang zeros are calculated,
including a specific regime that $\beta\gamma$ is very small, and the result reveals an interesting phenomenon that both parameters can 
reach their theoretical precision limit at the Lee-Yang zeros, and the probe qubit cannot work as a thermometer at a Lee-Yang zero if it 
sits on the unit circle. 

Apart from the Lee-Yang zeros and quantum Fisher information matrix, many other properties of the proposed nonlinear model are also 
worth studying, such as the existence of phase transitions or symmetries, and their connections with Lee-Yang zeros, the potential physical 
realizations of this model, and the generation and storage of spin squeezing with it. We believe that the further investigations of this model 
would help the community better understand the roles of nonlinearity in quantum spin models and its effect and potential usage in quantum 
information science, especially in quantum metrology. 

\begin{acknowledgments}
The authors would like to thank Dr. Mao Zhang, Dr. Zhucheng Zhang, and Dr. Lei Shao for helpful discussions, as well as two anonymous 
referees for their insightful views and suggestions. This work was supported by the National Natural Science Foundation of China (Grants 
No.\,12175075, No.\,11935012, and No.\,12247158). Y.G.S. also acknowledges the support from the ``Wuhan Talent'' (Outstanding Young 
Talents) and Postdoctoral Innovative Research Post in Hubei Province.

\end{acknowledgments}

\end{document}